\documentclass[%
aip,
amsmath,amssymb,
reprint,%
]{revtex4-2}
\usepackage{graphicx}
\usepackage{dcolumn}
\usepackage{calc}
\usepackage{bm}
\usepackage[utf8]{inputenc}
\usepackage[T1]{fontenc}
\usepackage{mathptmx}
\usepackage{etoolbox}
\usepackage{xcolor}
\usepackage{soul}
\usepackage[colorlinks=true,linkcolor=blue,citecolor=blue]{hyperref}%
\usepackage{amsfonts,amsmath,amssymb,amsthm,relsize}
\usepackage{hyperref}
\def\beq{\begin{equation}}
\def\eeq{\end{equation}}
\def\bea{\begin{eqnarray}}
\def\eea{\end{eqnarray}}

\makeatletter
\def\@email#1#2{%
\endgroup
\patchcmd{\titleblock@produce}
{\frontmatter@RRAPformat}
{\frontmatter@RRAPformat{\produce@RRAP{*#1\href{mailto:#2}{#2}}}\frontmatter@RRAPformat}
{}{}
}%
\makeatother
\newcommand{\kh}

\begin{document}

\preprint{AIP/123-QED}

\title{Complexity measure of extreme events}
	\author{Dhiman Das}
	\affiliation{Physics and Applied Mathematics Unit, Indian Statistical Institute, Kolkata 700108, India}
\author{Arnob Ray}
\affiliation{Physics and Applied Mathematics Unit, Indian Statistical Institute, Kolkata 700108, India}
 \affiliation{Artificial Intelligence for Climate and Sustainability, The Institute for Experiential Artificial Intelligence, Northeastern University, Portland 04101, ME, USA}
\author{Chittaranjan Hens}	
	\affiliation{Center for Computational Natural Science and Bioinformatics, International Institute of Informational Technology, Gachibowli, Hyderabad-500032, India}
	\author{Dibakar Ghosh*}
	\affiliation{Physics and Applied Mathematics Unit, Indian Statistical Institute, Kolkata 700108, India}	\email{dibakar@isical.ac.in}	
  \author{Md. Kamrul Hassan}
 \affiliation{Department of Physics, Dhaka University, Dhaka 1000, Bangladesh}
 \author{Artur Dabrowski}
 \affiliation{Division of Dynamics, Technical University of Lodz, Stefanovskiego 1/15, 90-924 Lodz, Poland}
 \author{Tomasz Kapitaniak*}
 \affiliation{Division of Dynamics, Technical University of Lodz, Stefanovskiego 1/15, 90-924 Lodz, Poland}
 \email{tomasz.kapitaniak@p.lodz.pl}
   
\author{Syamal K. Dana}	
\affiliation{Division of Dynamics, Technical University of Lodz, Stefanovskiego 1/15, 90-924 Lodz, Poland}
\affiliation{Centre for Mathematical Biology and Ecology, Department of Mathematics, Jadavpur University, Kolkata 700032, India}


\date{\today}

 \date{\today}
\begin{abstract}

 Complexity is an important metric for appropriate characterization of different classes of irregular signals, observed in the laboratory or in nature. The literature is already rich in the description of such measures using a variety of entropy and disequilibrium measures, separately or in combination. Chaotic signal was given prime importance in such studies while no such measure was proposed so far, how complex were the extreme events when compared to non-extreme chaos. We address  here this question of complexity in extreme events and investigate if we can distinguish them from non-extreme chaotic signal. The normalized Shannon entropy in combination with disequlibrium is used for our study and it is able to distinguish between extreme chaos and non-extreme chaos and moreover, it  depicts the transition points from periodic to extremes via Pomeau-Manneville intermittency and, from small amplitude to large amplitude chaos and its transition to extremes via interior crisis. We report a general trend of complexity against a system parameter that increases during a transition to extreme events, reaches a maximum, and then starts decreasing.  We employ three models, a nonautonomous Li\'enard system, 2-dimensional  Ikeda map and a 6-dimensional coupled Hindmarh-Rose system to validate our proposition.

\end{abstract}
\maketitle

\begin{quotation}

Complexity of  observed signal or time series is an important issue of concern for appropriate characterization since long. Past studies focused on the complexity of chaotic signals using Lyapunov exponents, fractal dimension, and a variety of entropies, including the permutation entropy, as the more recent one. It was possible to delineate periodic from  chaotic signals using such entropy measures. 
However, our interest is on extreme events that are not investigated so far for characterization of their complexity. 
We expect a change in complexity of extreme events that show occasional large events besides being chaotic. Many dynamical systems show  a transition from periodic to chaotic state (small amplitude) and then a sudden transition to large amplitude extreme events against a parameter via nonlinear processes, namely, Pomeau-Manneville (PM) intermittency and interior crisis to mention a few.  We use Shannon entropy in combination with disequilibrium to define complexity that can delineate nominal chaos (non-extreme) and extreme chaos and, their transitions. We use three different varieties of models, a non-autonomous system, a 2-dimensional map and a coupled system to validate our complexity measure. Results follow a similar trend for the nonlinear processes of origin of the extreme events whatever may be the system dynamics. 
\end{quotation}

\maketitle
\section{Introduction}
The knowledge of {\it extreme events}, particularly in deterministic dynamical systems, has advanced significantly during the past three decades\cite{chowdhury2022extreme, pal2023extreme, farazmand2019extreme, mishra2020routes, ghil2011extreme,sapsis2018new}. Recent studies have primarily focused on real-world extreme phenomena.  Extreme events are recognized as significantly large occasional deviations from long-term  nominal behavior. The large deviations are short-living but occur recurrently or non-recurrently. Such extremes often have a size of more than 4 to 8 times the standard deviation of mean event size \cite{reinoso2013extreme}. 
These phenomena have been studied across various disciplines, including oceanography \cite{kharif2003physical, akhmediev2016roadmap},  climate science \cite{ray2020understanding}, sociology, economics \cite{jusup2022social,helbing2015saving} and natural events like floods\ \cite{goswami2006increasing}, earthquake and  forest fire, for many years\cite{albeverio2006extreme}. Since then a number of studies have been  made on extreme events in a variety of systems including experiments such as optical fiber \cite{solli2007optical}, laser systems\ \cite{bonatto2011deterministic,zamora2013rogue, kingston2021instabilities, leo2023transition}, mechanical systems\ \cite{sudharsan2021emergence,meiyazhagan2021model}, electronic circuits\ \cite{cavalcante2013predictability,de2016local}, and other systems\ \cite{kumarasamy2018extreme,kumarasamy2022emergence}.
 To develop a better understanding of extreme events, in general, the dynamical models are often used as basic tools for studies in addition to dealing with real-world data. These efforts led to finding tools which  are now able to explain the origin of extremes in high-dimensional systems and networks of systems \cite{ray2020extreme, chowdhury2021extreme, ray2022extreme} and more complex natural events\ \cite{ghil2011extreme, farazmand2019extreme}. 
\par The primary concern of extreme events research is to predict the  time of arrival of such extremes and to address the most difficult question of predicting the magnitude of events. 
We  address here another important question how to measure complexity that may distinguish extreme events from non-extreme complex dynamics such as the nominal chaos and if possible, locate the transition point from non-extreme to extreme events  and vice versa in a parameter space. It is expected that complexity of extreme events shall be different due to the presence of occasional larger events, and  distinguishable from nominal chaos when the trajectory is bounded in state space. 

\par The existing complexity measures depend primary on three categories of metrics, namely, Lyapunov exponent (LE) \cite{wolf1985determining}, fractal dimension \cite{mandelbrot1985self, theiler1990estimating} and entropy \cite{shannon1948mathematical, cover1999elements} that  are usually used for  chaos in deterministic dynamical systems. Researchers from other disciplines, physiology \cite{goldberger2005genomic}, painting \cite{yang2003information}, and machine learning \cite{bialek2001predictability}, showed interest in identifying the complexity of signals from the general entropy measure. We emphasize here on Shannon entropy, which has been mainly proposed  \cite{shannon1948bell} as a measure of information content in a communication channel, however, it has also been used later for divergence measures \cite{lin1991divergence}. Several other measures \cite{shiner1999simple} have been developed from the Shannon entropy and used more frequently to characterize and for a possible quantification of complexity of dynamical systems' behaviors from time series data. It includes, but is not limited to, topological entropy that measures the exponential growth rate of the systems' distinguishable orbits \cite{adler1965topological}, Kolmogorov–Sinai entropy that measures chaos and complexity of motion that occurs in dynamical systems \cite{latora1999kolmogorov}. However, the Shannon entropy itself, based on symbolic dynamics \cite{kurths1995quantitative} or recurrence plot \cite{letellier2006estimating} although naturally produced large increase in complexity during a transition from periodic to chaotic state with a large amplitude, but they failed to recognize the emergence of extremes against a change of parameter.

On the other hand, permutation entropy (PE) was proposed \cite{bandt2002permutation} as an important complexity measure basing on identification of ordinal patterns and their probability in a data sequence. The PE has dependence on selective parameters, namely, the embedding dimension and time delay.   From a close look at the results presented earlier  \cite{bandt2002permutation, cao2004detecting}, it is noticed that PE shows a large change of complexity during a transition from periodic or small amplitude to large amplitude  chaos  at least in the logistic map. The complexity then slowly increases reaching a saturation against the system parameter. 
 Alternatively,  entropy and disequilibrium have been combined \cite{lopez1995statistical, martin2003statistical} for  statistical complexity measure of observed signals in many systems.
 
Disequilibrium is  based on  Euclidean distance \cite{lopez1995statistical},  Wootters distance \cite{wootters1981statistical}, entropy of Kullback-Shannon (or Tsallis or Rényi) \cite{aczel1975measures}, and the divergence of Jensen-Shannon (or Tsallis or Rényi) \cite{lamberti2004intensive}. 
In a recent study, PE in combination with  Jensen-Shannon disequilibrium was used \cite{xiong2020complexity} to measure complexity of chaos in logistic map.  This work was also successful to record both the sudden large change in complexity during transitions from low amplitude chaos to large amplitude chaos against a parameter and the transition from periodic to large amplitude chaos in another parameter range.  Once again, complexity increases sharply from small amplitude to large amplitude chaos and then saturates against a system parameter. A similar trend is seen during a transition from periodic to large amplitude chaos. However, none of the earlier works \cite{kurths1995quantitative,letellier2006estimating, bandt2002permutation, cao2004detecting,xiong2020complexity} dealt with systems or time series that showed occasional extremely large amplitude events. We are concerned about  signals that show occasional large events.
 \par How complex are the extreme events and it changes during transitions from periodic to large amplitude chaotic state via PM intermittency \cite{manneville1980different} and low amplitude chaos to large amplitude chaos via interior crisis \cite{grebogi1987critical} when signature of extreme events does appear? Three exemplary systems that show origin of extreme events, are used for numerical validation of our proposed measure: Forced Li{\'e}nard system \cite{kingston2017extreme}, the Ikeda map \cite{ray2019intermittent} and a 6-dimensional coupled spiking-bursting Hindmarsh-Rose (HR) systems \cite{hindmarsh1984model, mishra2018dragon} that shows a more complex basin structure \cite{storace2008hindmarsh}. We consider the product of Shannon entropy and disequilibrium that provides a consistency in complexity measure. The Shannon entropy is preferred to PE since it records information of each local maxima of a time series or a data sequence and then takes care of their probability measures. In PE measure, information on occasional large events are missing during symbolic representation of data sequence in search of ordinal patterns.
\par We plot the bifurcation diagram  for each of the considered systems against a  parameter and delineate the regions of transitions via PM intermittency and interior crisis when extreme events appear. And capture the corresponding signal or time series (long enough) for each parameter and take care of each events (local maxima) in a time series or data sequence so that no information on the occasional large events are missed or ignored. Then we derive the normalized Shannon entropy and disequlibrium to estimate complexity as a product of both. {We plot the complexity measure against a parameter and compare with the bifurcation diagram of each of the systems. 
The complexity shows a sharp increase during the PM intermittency when a transition occurs from periodicity to large amplitude chaos with the presence of occasional extreme events. After reaching a maximum against a parameter, complexity continues to decrease slowly with a  transition to non-extreme chaos. Similarly, during the interior-crisis, we find a sharp rise in complexity during the transition from small amplitude chaos to large amplitude chaos with occasional extremes and then a slow decrease in complexity against a parameter after reaching a maximum. Our results show a trend of complexity different from the earlier reports \cite{bandt2002permutation, cao2004detecting, lopez1995statistical, martin2003statistical, xiong2020complexity}.
and justify a trend of slow decrease in complexity beyond the regime of extreme events when the occasional large events gradually become more and more frequent. 
 Thereby we are able to distinguish non-extreme chaotic events and extreme events and, their transitions against a parameter. 
 For all the three systems, a similar trend is found for the two different nonlinear processes of origin of extreme events.



\section{Complexity Measure}\label{methodology}
 We describe here Shannon entropy and disequilibrium distance, which are frequently used for physical systems and for a wide variety of applications\cite{kowalski2011distances}, and then define our proposed complexity measure. 
Let us first consider a time series consisting of a sequence of $n$ datasets, say $\{x_{1}, x_{2},\cdots,x_{n}\}$, where  $x_n$ is the peak or local maxima of a data sequence or time series. The data sequence is divided into $m$ number of bins, and the width or length of each bin is estimated from the difference between the maximum and minimum values in the data sequence and then by dividing this difference by $m$. The  corresponding probabilities $\{p_1, p_2,\cdots,p_m\}$ of all the  bins are estimated so that $\sum_{i=1}^m p_i =1$. The {\it information entropy}~ \cite{shannon1948mathematical},  in other words, the Shannon entropy $H$ of the given time series is then estimated, \begin{equation*}
    H=-\sum_{i=1}^m p_i~\log_2 p_i. 
\end{equation*}
For a normalized version of the information entropy $H$, we calculate the maximum information entropy as follows:
$H_{max} = log_2 m$ where
$p_e=\{1/m,1/m,1/m,...,1/m\}$ is the set of probabilities of a uniform distribution. 
The normalized Shannon entropy $ S $ is \begin{equation}\label{eqn_1}
     S=H/H_{max}. 
\end{equation}
 If the value of $S$ is zero, it represents an ordered state. 
 For the case of $S>0$, we get the information that the time series represent an increasing disorder state. 
  Examples are given in Appendix A to elaborate how the maximum and minimum values of entropy are chosen and the bin width is selected and then how the Shannon entropy is calculated considering the information of a complex signal of varying amplitude; no information on occasional large events are missed out.




\par Next we define $D$ as {\it disequilibrium distance}~\cite{nicolis1977self, kowalski2011distances,lopez1995statistical}, between two probability distributions. Specifically, it measures  the distance of a distribution of an observable quantity from a uniform distribution. 
Two conditions are imposed on the maximum limit of {\it disequilibrium distance}: $D>0$ to have a positive measure of complexity and $D=0$ for the equi-probability limit. Note that for period-1 dynamics $D=1$. A larger positive value of $D$ indicates that the probability distribution is highly deviating from the uniform distribution. The Euclidean distance between two sets of probabilities is then quantified,
\[\sum_{i=1}^m \bigg(p_i-\frac{1}{m}\bigg)^2, 
\]
when the normalized disequilibrium distance is given by 
\begin{equation}
D = \frac{m}{m-1} \sum_{i=1}^m \bigg(p_i-\frac{1}{m}\bigg)^2.
\end{equation}

\par Finally, we introduce the {\it complexity measure}\cite{lopez1995statistical} as denoted by $C$ for a sequence of $m$ numbers of probabilities. This combines the dual measures  of the Shannon entropy $S$ and disequilibrium distance $D$ of a time series to define the probability-based complexity measure, 
\begin{equation}
    C=\left(-\frac{1}{\log_2 m}\sum_{i=1}^m p_i \log_2p_i\right)\left(\frac{m}{m-1} \sum_{i=1}^m (p_i-1/m)^2\right). 
\end{equation}
$C$ plays a distinctive roles in distinguishing extreme events from non-extremes and there transition from one to the other, as elaborated with examples in Section III.
\section{Results}
\label{result}
Results of three dynamical systems, namely, a Li{\'e}nard type oscillatory system, the Ikeda map and a coupled spiking-bursting Hindmarsh-Rose model are presented. They exhibit onset of extreme events in selective ranges of parameters following two different nonlinear dynamical processes.  A significant height measure  is used as a referral marker for identification of extreme events; however, it is not used for estimation of complexity measure.
To  estimate the normalized Shannon entropy~($S$), Euclidean distance~($D$) and then complexity ($C$), we collect local maxima of numerically generated time series and the number of bins $m=50$ is set for all the examples. The effect of  varying number of bins on $C$  has been checked for the three systems as presented in Fig.~\ref{fig5} of Appendix B that the transition points are robust to changes in the number of bins during both the nonlinear processes, PM intermittency and interior crisis induced extreme events.  However, a change in the quantitative values of $C$ is observed: it provides a qualitative picture of complexity in a data sequence or time series; no unique quantitative measure is possible so far.
\par  We define the   significant height measure $T=\mu \pm d\sigma$ where $\mu$ is the mean height of a time series, $\sigma$ is the standard deviation and $d$ is chosen arbitrarily from range of values $4-8$. If the amplitude of an event (local maxima or minima)  is larger or smaller than $T$, it is marked as extreme. By this referral marker, we made a tentative identification or an intuitive idea of extreme events in a long time series and the range of parameters where extreme events may originate in dynamical systems. The complexity measure is independent of the significant height $T$. 
\subsection{Forced Li{\'e}nard system}
A Li{\'e}nard type oscillator with a periodic forcing is 
\begin{equation} \label{eqn.4}
	\begin{array}{lcl}
     \dot{x}=y,\\
	\dot{y}=-\alpha x y-\gamma x-\beta x^3+\uppercase{F} \sin{\omega t},
\end{array}
\end{equation}
where $\alpha$ and $\beta$ are the nonlinear damping and the strength of non-linearity, respectively. $\gamma$ is related to the internal frequency of the autonomous system, $F$ and  $\omega$ are the amplitude and frequency of an external sinusoidal signal, respectively. The parameters are selected as $\alpha=0.45$, $\beta=0.50$, $\gamma=-0.50$ and $F=0.2$ for our numerical simulations.

\begin{figure}
\centering
\includegraphics[scale=0.50]{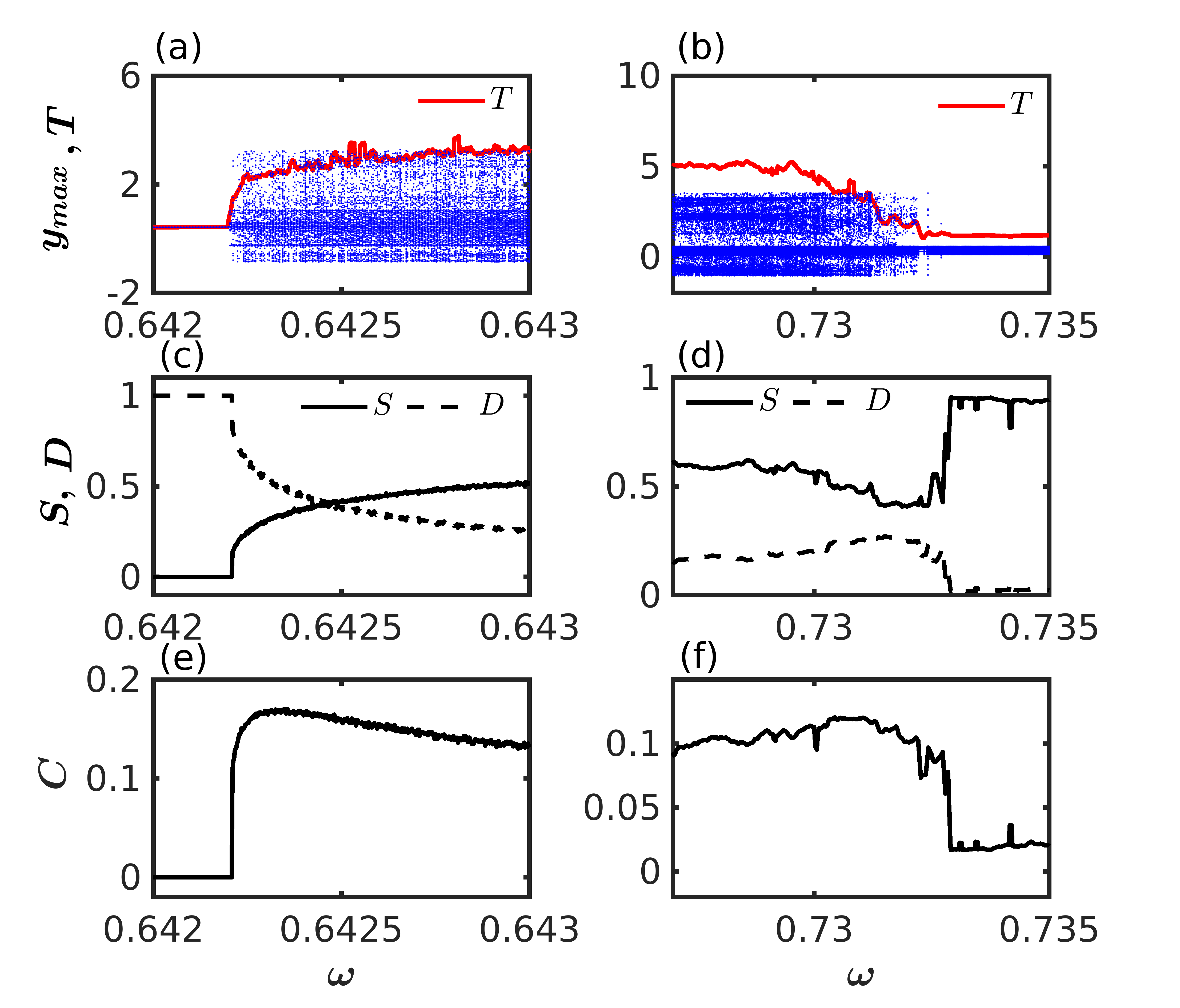}
\caption{\textbf{Forced Li\'enard system}: Bifurcation diagrams of $y_{\max}$ against forcing frequency $\omega$. (a) PM intermittency: Periodic oscillation transition to large amplitude events at a critical $\omega \approx 0.642212$ and it continues for larger $\omega$. (b) Interior-crisis-induced extreme events: Small amplitude chaos transition to large amplitude chaos that  continues to appear for decreasing $\omega$. Normalized Shannon entropy $S$ (solid black line) and disequilibrium distance $D$ (dashed black line) in the PM intermittency regime (c) and interior-crisis region (d). Behavior of complexity $C$ against $\omega$ are shown for the PM intermittency region (e) and interior crisis route (f).  Number of bins $m=50$, for both cases, other parameters are $\alpha=0.45$, $\beta=0.50$, and $\gamma=-0.50$. A significant height threshold $T=\mu +6\sigma$ is used as a referral marker of extreme events.}
\label{fig1}
\end{figure}
The forced Li{\'e}nard system exhibits a wide range of dynamics against the forcing frequency $\omega$. A plot of local maxima $y_{max}$ in the range $\omega \in [0.642,0.643]$ is shown in Fig.~\ref{fig1}(a). At a critical value of $\omega \approx  0.642212$, an abrupt change from period-1 limit cycle to a large amplitude oscillation is noticed. Earlier, Kingston et al.\cite{kingston2017extreme} have demonstrated the onset of extreme events via PM intermittency at this critical $\omega$ value. The large amplitude oscillation  exhibits a nearly periodic oscillation (laminar phase) interrupted by intermittent large amplitude chaotic bursts (turbulent phase). And this intermittent chaotic bursting continues for a range of $\omega$ values when large bursting events slowly become more and more frequent, leading to large amplitude non-extreme chaos. A significant height line $T=\mu+6\sigma$  is drawn (red line) against $\omega$. Some of the occasional chaotic bursts in the turbulent phase have heights ($y_{max}$) larger  than the $T$ line (red line) and marked as extremes in Fig.~\ref{fig1}(a). 
A similar bifurcation plot of $y_{max}$ is drawn in Fig.~\ref{fig1}(b), for a range of $\omega \in [0.727,0.735]$, where another large change in $y_{max}$ erupts at $\omega \approx 0.73242$. This large and sudden change in $y_{max}$ occurs from low amplitude chaos at this critical point via interior crisis\cite{kingston2017extreme, mishra2020routes} and it also continues for a range of $\omega$. The large amplitude oscillations are once again identified as extreme events  by the marker $T$ (red line) for a specified range of $\omega$. The extremely large events slowly become more frequent with decreasing $\omega$, leading to large amplitude non-extreme chaos when the bifurcation diagram Fig.~\ref{fig1}(b) appears more dense. 
 The normalized Shannon entropy $S$ measures disorder or randomness in a time series, whereas $D$ indicates how much it deviates  from the uniform distribution, in other words, from the long time mean value. In fact, they are showing opposite trends against $\omega$ in Fig.~\ref{fig1}(c) that corresponds to Fig.~\ref{fig1}(a). 
For periodic oscillation, the probability of occurrence of an event is always $1$ due to a constant peak value in a data sequence and hence $D$ (dashed black line) is $1$ while the normalized Shannon entropy (solid black line) is $0$ as shown in Fig.~\ref{fig1}(c) for PM intermittency. The intermittent bursting starts appearing at $\omega \approx 0.642212$ that we define as extreme events and, we find an increasing entropy $S$ against $\omega$ as expected due to the origin of instability in the form of intermittent bursting. 
While $D$ starts at $1$ for periodic oscillation and suddenly drops at the transition point with the onset of intermittency, i.e., extreme events as defined by our maker $T$ and follows a decreasing trend against $\omega$. On the other hand, during the origin of instability in low amplitude chaos via interior crisis in Fig.~\ref{fig1}(d), $S$ decreases suddenly when small amplitude chaos transition to extreme events what is unexpected and confusing. However, as expected $D$ increases against $\omega$ when divergence from uniform distribution in chaotic sequence increases with the onset of extreme events. For two different nonlinear processes, i.e., onset of PM intermittency and interior crisis, we find contradictory results from the normalized entropy $S$ regarding increasing disorder due to emergence of extreme events. However, when we combine $S$ and $D$ to define complexity $C$ for the two nonlinear processes as shown in Figs. \ref{fig1}(e) and \ref{fig1}(f), results are consistent for both the processes. The Shannon entropy $S$  alone fails to provide consistency in the complexity measure of extreme events. 

\par Figures~\ref{fig1}(e) and \ref{fig1}(f) depict the  complexity measure $C$ for varying $\omega$ during the PM intermittency and interior-crisis, respectively. The value of $C$ is $0$  for the period-1 oscillation at left in Fig.~\ref{fig1}(e). At $\omega\approx 0.642212$, the system transition from periodic to  extreme events via PM intermittency, when $C$ abruptly rises. 
Then $C$ increases slowly to reach a maximum  and finally starts decreasing against $\omega$, when the system dynamics gradually changes from extreme to non-extreme chaotic oscillation. No sharp transition from extreme to non-extreme chaos is seen. 
Figure \ref{fig1}(f) shows a similar trend in $C$ for decreasing $\omega$ starting from the right side. The small amplitude chaotic oscillation transition to large amplitude extreme  events at $\omega\approx 0.73242$; there is a significant jump in $C$. First an increase to a maximum and then a slow decreasing trend of $C$ is observed against a decreasing $\omega$ when extreme events transition to non-extreme chaotic oscillation. In fact, more and more frequent large events start appearing with decreasing $\omega$, leading to large amplitude non-extreme chaos. 


 \begin{figure}[hpt]
\centering
\includegraphics[scale=0.5]{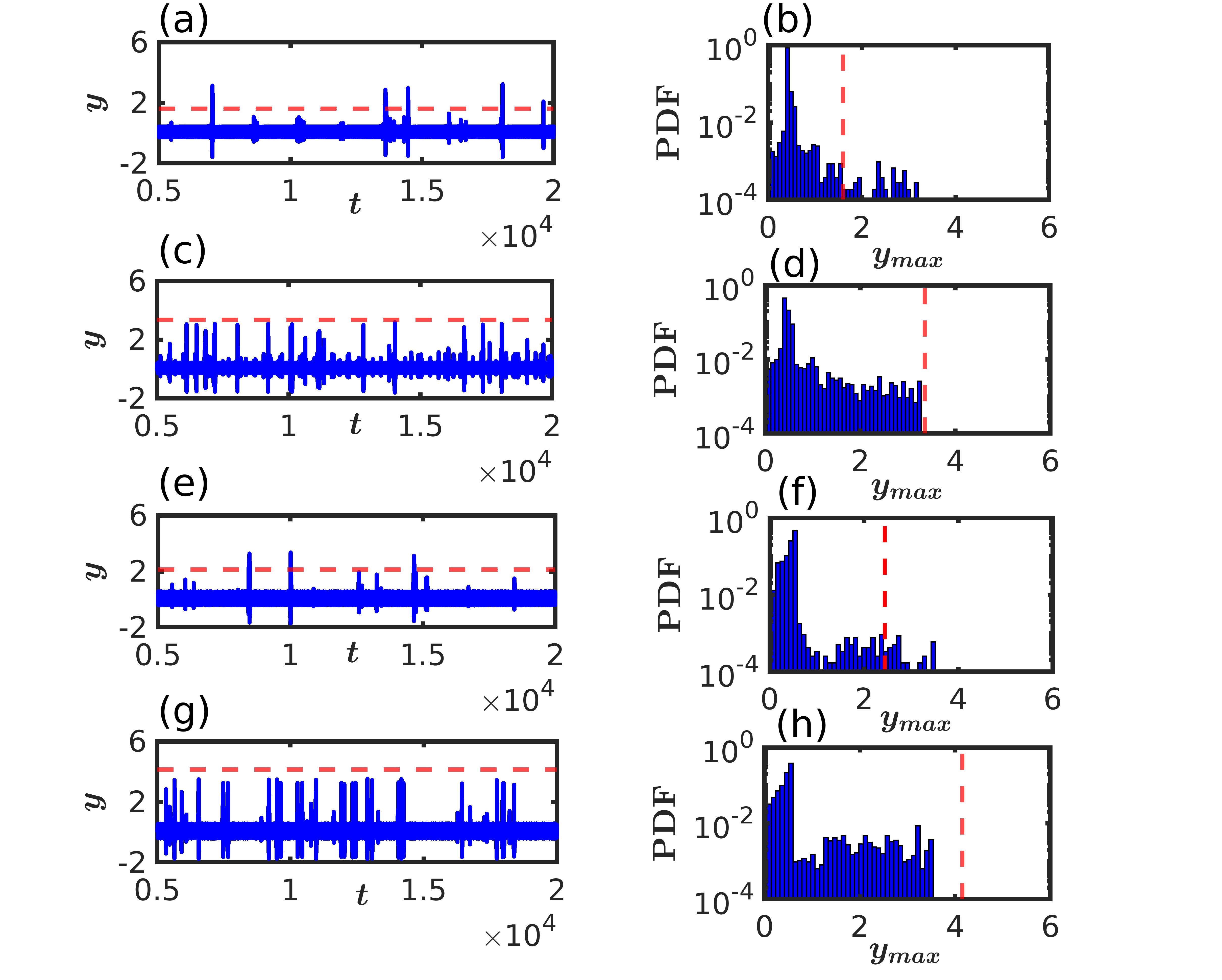}
\caption{\textbf{Forced Li\'enard system}. PM intermittency: Temporal dynamics of extremes for (a) $\omega$= $0.642236$, non-extreme events for (c) $\omega$= $0.643$. PDF in semi-log scale for (b) $\omega$=0.642236 and (d) $\omega=0.643$ during PM intermittency.
Interior crisis: Temporal dynamics of extreme events for (e) $\omega=0.731693$  and non-extremes for (g) $\omega$= $0.73007$. PDF in semi-log scale  for (f) $\omega=0.731693$  and (h) $\omega=0.73007$. Extreme events are larger than the significant height (horizontal dashed red lines) in (a) and (e) when their PDF of events (b) and (f), respectively, show  rare occurrence of large events beyond the vertical dashed marker line. Non-extreme chaotic events are very frequent in (c) and (g) when all the events are of lower height than the significant height and show no events beyond the marker $T$ (vertical dashed line) in (d) and (h), respectively. 
}
\label{fig2}
\end{figure}
    
\par Now we confirm the existence of extreme events and distinguish them from non-extreme chaotic events in Fig.~\ref{fig2} as claimed in Figs.~\ref{fig1}(a, b) by plotting the temporal dynamics of $y$ variable and a corresponding probability density function (PDF) plot of $y_{max}$ in semi-log scale for four different values of $\omega$. First of all, we consider the origin of extremes via PM intermittency: The time evolution of $y$ in Fig.~\ref{fig2}(a) shows extreme events for $\omega=0.642236$ since the occasional large peaks are larger than the marker $T$ line (horizontal dashed red line). The  PDF against the size  of events confirms rare occurrence of large events larger than the vertical $T$ line in Fig.~\ref{fig2}(b). Figure~\ref{fig2}(c) shows the time evolution of $y$ showing more frequent large  events  for $\omega=0.643$. The frequent large events are of lower height than the $T$ (horizontal dashed red line). The corresponding PDF in Fig.~\ref{fig2}(d) shows almost no events beyond the vertical $T$ line (vertical red dashed line). 
While for $\omega= 0.731693$, the time series in Fig.~\ref{fig2}(e) and their PDF in Fig.~\ref{fig2}(f) confirm occurrence of occasional large events during interior-crisis.  
 Figure~2(g, h) present the time evolution and corresponding PDF  when a slow transition to non-extremes occurs with more frequent events against a decreasing $\omega$. Large events becomes more frequent for $\omega= 0.73007$ when the PDF shows no events beyond the $T$ line (vertical dashed red line) showing a slow transition to non-extreme chaotic events. 

    \begin{figure}[hpt]
    \centering
    \includegraphics[scale=0.5]{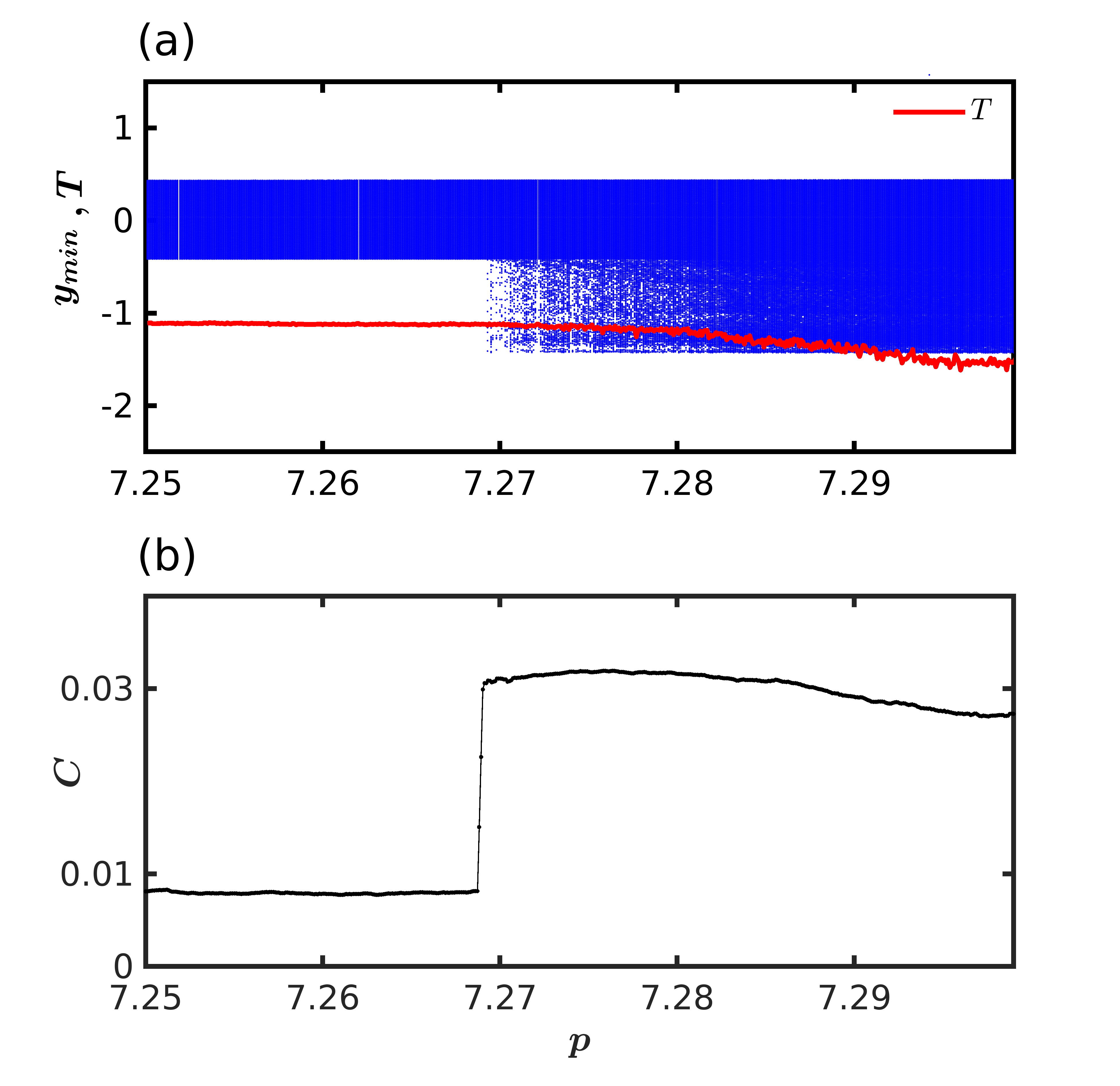}
\caption{\textbf{Ikeda map}. (a) Bifurcation diagram of $y_{min}$ and significant height $T$ (solid red line) against $p$. Sudden expansion of the attractor at a critical point $p \approx 7.26884894$ via interior crisis. 
(b) Complexity measure $C$ (black solid line) against $p$. $C$ remains low during non-extreme chaotic oscillation until it transitions to a high value at the critical point $p \approx 7.26884894$ with the onset of extremes. 
Parameters $A = 0.85, B = 0.9$ and $k = 0.4$.}
    \label{fig3}
    \end{figure}
\subsection{Ikeda map}
Our second example is a simplified version of the Ikeda map,
\begin{equation} \label{eqn.5}
\begin{array}{lcl}
z_{n+1} = A + B z_n \exp\left[ik - \frac{ip}{1 + |z_{n}|^2}\right],
\end{array}
\end{equation}
where $z_{n}=x_{n}+i y_{n}$. The evolution of  laser over a nonlinear optical resonator is described by this model. 
The two-dimensional Ikeda map with real values is obtained as \\
\begin{equation}
\begin{aligned}
    x_{n+1} &= A + B x_{n} \cos\left(k - \frac{p}{w_{n}}\right) - B y_{n} \sin\left(k - \frac{p}{w_{n}}\right), \\
    y_{n+1} &= B y_{n} \cos\left(k - \frac{p}{w_{n}}\right) + B x_{n} \sin\left(k - \frac{p}{w_{n}}\right),
\end{aligned}
\end{equation}
where $A$ and $B$ represents  the laser input amplitude and  the coefficient of reflectivity of the partially
reflecting mirrors of the cavity. $w_{n}=1+x_{n}^2+y_{n}^2$, and $k$ is the laser-empty-cavity detuning,
and $p$ measures the detuning due to the presence of a non-linear medium in the cavity. We select the parameters $A=0.85$, $B=0.9$ and $k=0.4$ for our numerical simulations and take $p$ as a control parameter. 
The Ikeda map demonstrates the origin of chaos for a range of $p$ values, bounded within a state space. When $p>p_c$, a critical value, the trajectory occasionally travels to distant regions in the state space far away from the bounded limit, due to interior crisis \cite{grebogi1987critical}, but returns to the bounded region after a while. The temporal dynamics show low amplitude chaos for $p<p_c$, but at the critical $p_c$, starts  showing intermittent large amplitude spikes that represent the occasional faraway journey of the trajectory from the bounded region of the state space of the system. The infrequent large amplitude spikes suggest the appearance of extreme events \cite{ray2019intermittent}.
 
  \begin{figure}[hpt]
    \centering
   \includegraphics[scale=0.465]{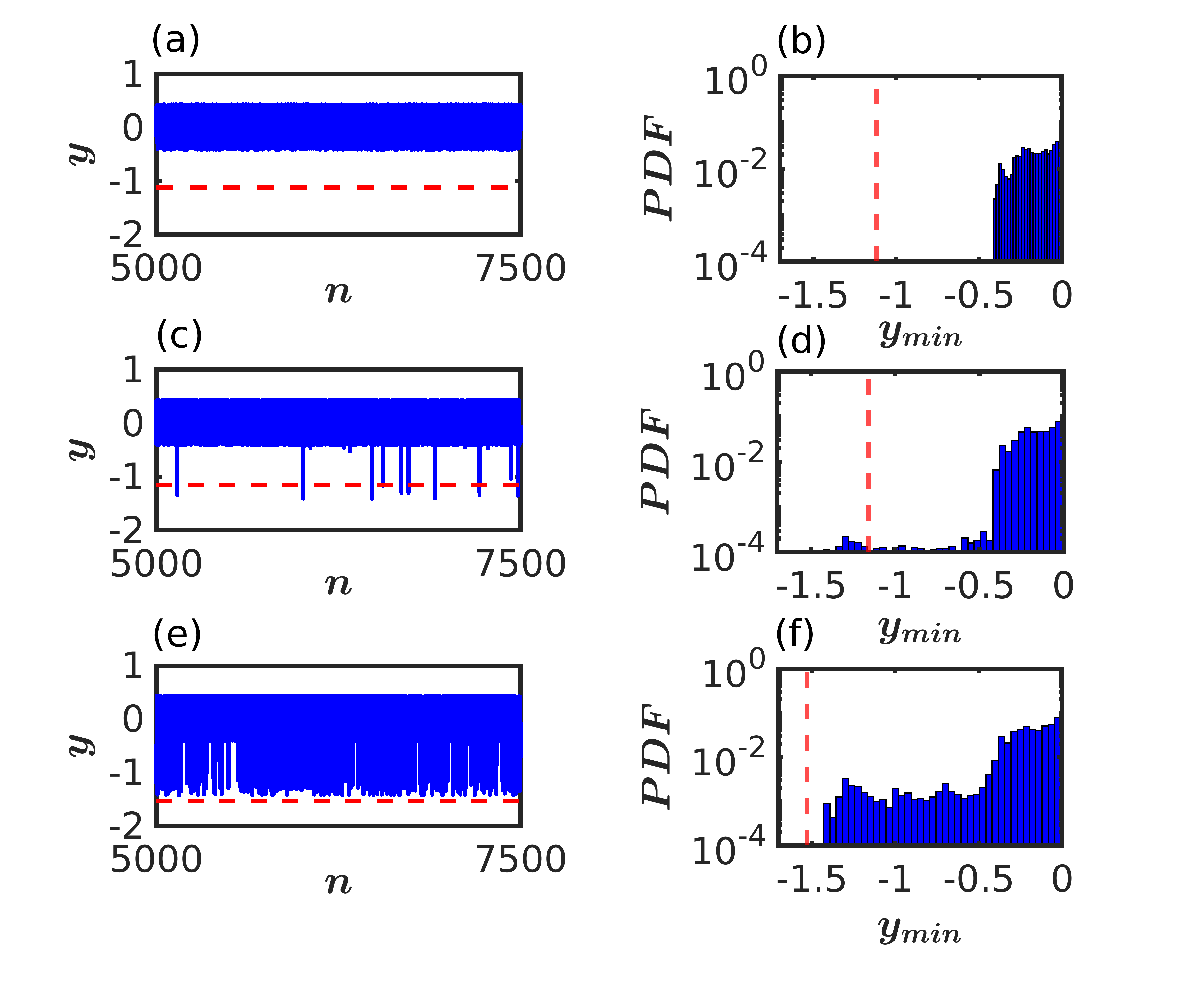}
\caption{\textbf{Ikeda map}. Temporal dynamics of $y$ in the pre-crisis non-extreme bounded chaos regime for $p = 7.265$ (a), post-crisis extremes for $p=7.275$ (c), and post-crisis frequent non-extreme large events for $p = 7.295$(e). PDF for pre-crisis region (b), extreme events (d) and post-crisis frequent large events (f). }
    \label{fig4}
    \end{figure}
\par As illustrated in a bifurcation diagram against $p$ in Fig.~\ref{fig3}(a), the map exhibits bounded chaos  for a range of $p$ and then a sudden large expansion, however, a negative increase in $y_{min}$ is seen at a critical $p=p_c$.  For a range of $p>p_c$, we mark appearance of extreme events by the marker $T=\mu-5\sigma$  (red line), where the large peaks $y_{min}$ cross the threshold. 
The negative sign in $T$ is chosen since we collect the local minima $y_{min}$. 
Figure~\ref{fig3}(b) shows the plot of complexity $C$ against $p$.  At $p=p_c \approx 7.26884894$, a sudden switching to a high value of $y_{min}$ is seen when small amplitude  chaos transition to occasional large amplitude extreme events via interior crisis. The complexity $C$ remains almost constant for a range $p$ values and then start decreasing. This slow decrease in $C$ indicates a gradual increase in the number of large spiking events against $p$ indicating appearance of more frequent large events. 
For a larger $p=7.295$, the temporal dynamics in Fig.~\ref{fig4}(e) shows very frequent large events when $C$ value starts decreasing. Accordingly, the PDF in Fig.~\ref{fig4}(f) confirms this behavior of the temporal dynamics with increased probability of events $y_{min}$ of larger size.

\subsection{Coupled Hindmarsh-Rose neuron model}
 A 6-dimensional system is formed using two coupled HR neuron models via synaptic chemical communication, 
\begin{align}
\dot{x}_i &= y_i + b x_i^2 - a x_i^3 - z_i + I - \epsilon (x_i-v_s) \Gamma(x_j),\\
\dot{y}_i &= c - d x_i^2 - y_i, \\
\dot{z}_i &= r[s(x_i - x_R) - z_i],
\end{align}
where \(i, j = 1, 2\) and \(i \neq j\) denote two oscillators. The chemical synaptic communication is represented \cite{mishra2018dragon}
by a sigmoidal function $\Gamma(x)$, 
\begin{equation}
\Gamma(x) = \frac{1}{1 + \exp^{- \lambda (x - \theta)}}
\end{equation}
The following parameters remain constant: $a = 1, b = 3, c = 1, d = 5, x_R = -1.6, r = 0.01, s = 5,I = 4$,
when the isolated neurons show periodic bursting. The parameters of the coupling function are chosen as $ v_s = 2.0,\lambda = 10.0 , \Theta = -0.25 $. The coupling strength $\epsilon$ is assumed identical and positive that represents a repulsive coupling (inhibitory interactions between the neurons).  
   \begin{figure}[hpt]
      \centering
\includegraphics[scale=0.475]{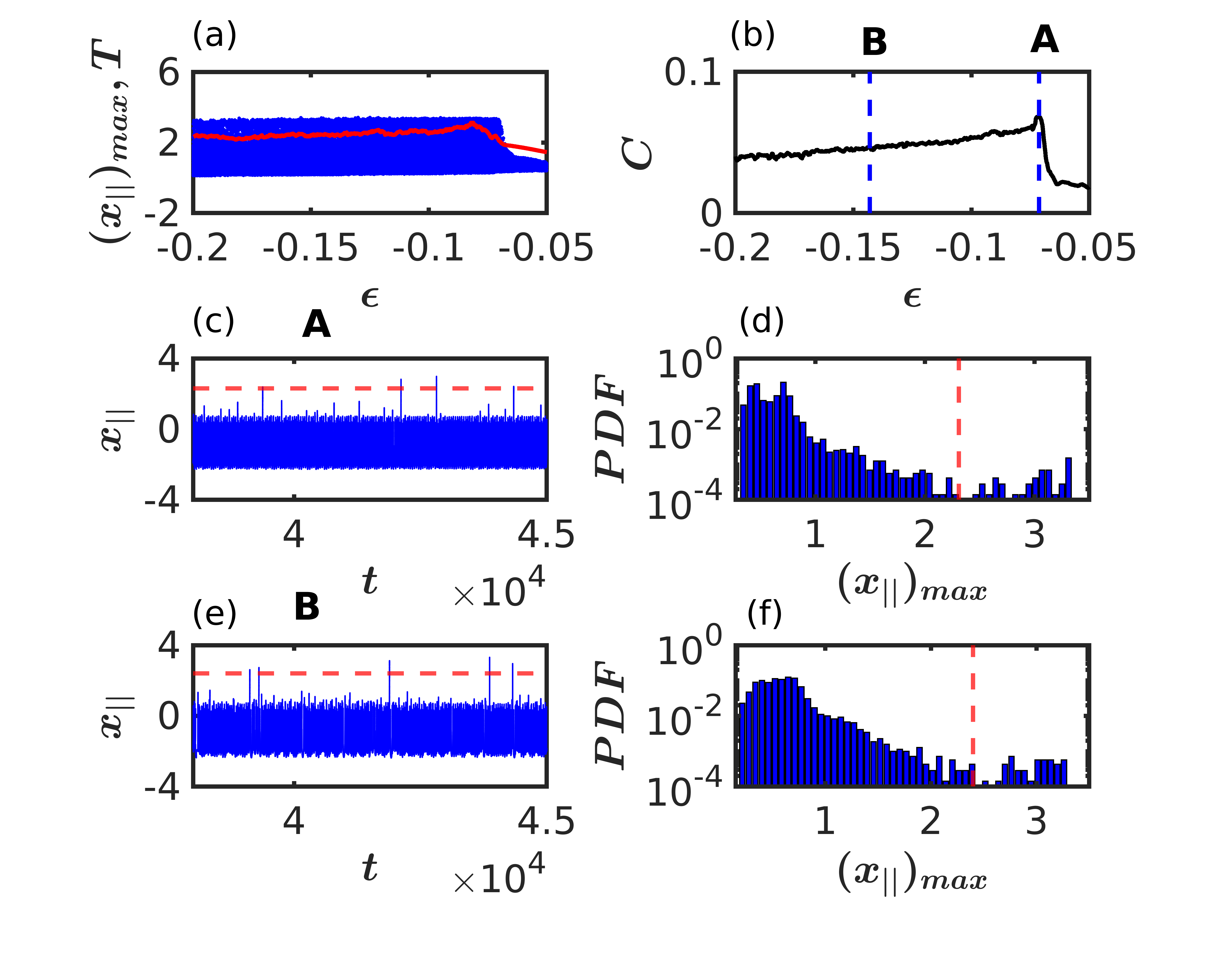}
\caption{\textbf{Coupled Hindmarsh-Rose neuron model}: (a) Bifurcation diagram of $x_{||}$ (blue dots) and extreme event marker $T$ (red line) against $\epsilon$. (b) Complexity measure $C$ (black line) of the coupled HR model against $\epsilon$. (c, e) Temporal dynamics of $x_{||}$ for $\epsilon = - 0.07127$ (A) and, $- 0.1431$ (B) when their corresponding distribution (PDF) of events are in  (d) and (f), respectively.}
    \label{fig5}
    \end{figure}
\par We create a bifurcation diagram of $x_{||_{max}}$ against the coupling parameter $\epsilon$ in 
 Fig.~5(a), where $x_{||} = x_1 + x_2 $ represents the antiphase manifold.  Instability in this manifold leads to the formation of quasiperiodic motion followed by chaos against the coupling strength.  Then the chaotic motion transitions to occasional large amplitude events via interior crisis at a critical $\epsilon$. As a result, infrequent  large amplitude chaotic bursting \cite{mishra2018dragon, mishra2020routes} starts occurring  that represents occasional bubbling from the antiphase manifold and the origin of extreme events. We capture the time evolution of $x_{||}$ for each coupling constant $\epsilon$ and then estimate $C$ for each time series. As mentioned above, the dynamics of $x_{||}$ first becomes quasiperiodic, followed by chaos and then a sudden jump in amplitude to originate occasional extreme events \cite{mishra2020routes}. The complexity $C$ follows the dynamical processes to show a gradual increase reaching a maximum during the origin of extreme events and then shows a slow decrease.
The time evolution of $x_{||}$ at two points (A, B) in Fig.~5(b) are shown in Fig.~5(c) and Fig.~5(e), respectively. The corresponding PDFs in Fig.~5(d, f) show that the distribution of occasional large events are more and more frequent with decreasing $\epsilon$.
\section{Conclusions}\label{summary}
 We proposed a complexity measure  to distinguish between non-extreme chaotic events and extreme events.  Earlier studies distinguished periodic and chaotic events and also transitions from low amplitude to large amplitude chaos using either permutation entropy \cite{bandt2002permutation, cao2004detecting} or a complexity measure \cite{xiong2020complexity} that uses a combination of permutation entropy and disequilibrium. None of the them focused on the problem of complexity  of occasional large events, i.e., extreme events and how it differs from non-extreme chaotic events. We addressed this aspect of complexity using a combination of Shannon entropy and disequilibrium measures. The Shannon entropy is preferred to permutation entropy that loses information on occasional large events in course of symbolization of events in search of ordinal patterns in a data sequence, and their probability measure. On the other hand, Shannon entropy includes all peak events (local maxima) in a data sequence and estimates the probability of all size of events in a bin length. Thereby we avoided any loss of information of the occasional large events in a data sequence generated from our simulations in three paradigmatic models, the forced Li\'enard oscillator, the Ikeda map and a 6-dimensional coupled HR model.
 \par  Especially, we checked the validity of our results in the 6-dimensional coupled HR model that has a more complex basin structure \cite{storace2008hindmarsh}. Two nonlinear processes, namely, PM intermittency and interior crisis were focused that show transition to extreme events, either from periodic or quasiperiodic and small amplitude chaotic oscillations.
Our proposed complexity measure shows an increase during the onset of extreme events reaching a maximum against a parameter variation, during either of the nonlinear processes, and then follows a decreasing trend when the occasional large events are more and more frequent. And this trend of complexity against a system parameter is different from the trends in change of complexity in earlier studies \cite{bandt2002permutation, cao2004detecting, xiong2020complexity} where it follows an increasing trend and then saturates with the origin of large amplitude chaos (non-extreme). Our proposed  complexity measure shows a generic property independent of the three model systems considered here and follows a common trend for both the nonlinear processes involved in the origin of extreme events. It is expected to follow the same trend for other systems, including the memristor-based HR model \cite{xu2022modeling}. Note that the complexity measure provided qualitative information on the complexity of extreme events, and it is clearly larger than small amplitude or large amplitude chaos. Any unique quantitative measure of complexity is yet to be proposed, to the best of our knowledge.

\section*{Appendix A}
\label{appendix a}
The Shannon entropy \( H(X) \) of a discrete random variable \( X \) with probability distribution \( P(x_i) \) is given by

\[
H(X) = -\sum_{i} P(x_i) \log_2 P(x_i).
\]

To normalize the entropy, we divide by the logarithm of the number of bins $(\text{num\_bins})$,
\[
S=H_{\text{normalized}}(X) = \frac{H(X)}{\log_2 (\text{num\_bins})}.
\]

\subsection*{Example 1: Maximum Entropy}

Consider the dataset \( x = \{1, 2, 3, 4, 5, 6, 7\} \). Each value is equally likely.

\begin{itemize}
    \item Number of bins: 7
    \item Probability for each bin: \( P(x_i) = \frac{1}{7} \)
\end{itemize}

The entropy \( H(x) \) is calculated as follows,

\[
H(x) = - \sum_{i=1}^{7} P(x_i) \log_2 P(x_i).
\]

Since \( P(x_i) = \frac{1}{7} \), so

\[
H(x) = - \sum_{i=1}^{7} \frac{1}{7} \log_2 \frac{1}{7} = - 7 \times \frac{1}{7} \log_2 \frac{1}{7} = \log_2 7 \approx 2.807.
\]

The normalized entropy is

\[
H_{\text{normalized}}(x) = \frac{H(x)}{\log_2 7} = \frac{\log_2 7}{\log_2 7} = 1.
\]
This represents a most disordered state as clearly seen from the data sequence. 
\subsection*{Example 2: Minimum Entropy}

Consider the dataset \( x = \{5, 5, 5, 5, 5, 5, 5\} \). All values are the same.

\begin{itemize}
    \item Number of bins: 7
    \item Probability for the bin containing value 5: \( P(5) = 1 \)
    \item Probability for all other bins: \( P(x_i) = 0 \) for \( x_i \neq 5 \)
\end{itemize}

The entropy \( H(x) \) is calculated as follows,

\[
H(x) = - P(5) \log_2 P(5) - \sum_{i \neq 5} P(x_i) \log_2 P(x_i).
\]

Since \( P(5) = 1 \) and \( \log_2 1 = 0 \),

\[
H(x) = - 1 \times \log_2 1 - \sum_{i \neq 5} 0 \times \log_2 0 = 0.
\]

The normalized entropy is

\[
H_{\text{normalized}}(x) = \frac{H(x)}{\log_2 7} = \frac{0}{\log_2 7} = 0,
\]
when the time series maintain an order state.

 A data sequence is created by taking the local maxima or minima of a numerically generated  signal. They are stored in different numbers of bins. The length or width of a bin is decided from the difference of maximum and minimum values in the data sequence and then dividing the difference by the number of bins. Thus each bin has a width that consists of a data length representing the range of heights (amplitude) of a signal or data sequence. The bins are arranged in increasing order of height of a data sequence. Then the probability of a data set within  a range of height (length of a bin) in a bin is considered. The Shannon entropy is estimated by calculating the probabilities of data sets in all the bins. Obviously, the occasional large events within a range of height are kept in a particular bin having low probability; however, information of such large events is included in the entropy measure. This is the major difference of our proposed measure from the PE.
  \section*{Appendix B}
\label{appendix b}
We check the results for different bin sizes for all three systems. From Fig.~\ref{fig5}, we observe a similar pattern, no qualitative change in the basic features of $S$, $D$ and $C$ for different bin sizes 25, 50, 75, and 100. Basically, the transition points for all the systems do not change for both the nonlinear processes to extremes although finite values have changed.\\
\begin{figure}[hpt]
    \centering
\includegraphics[scale=0.325]{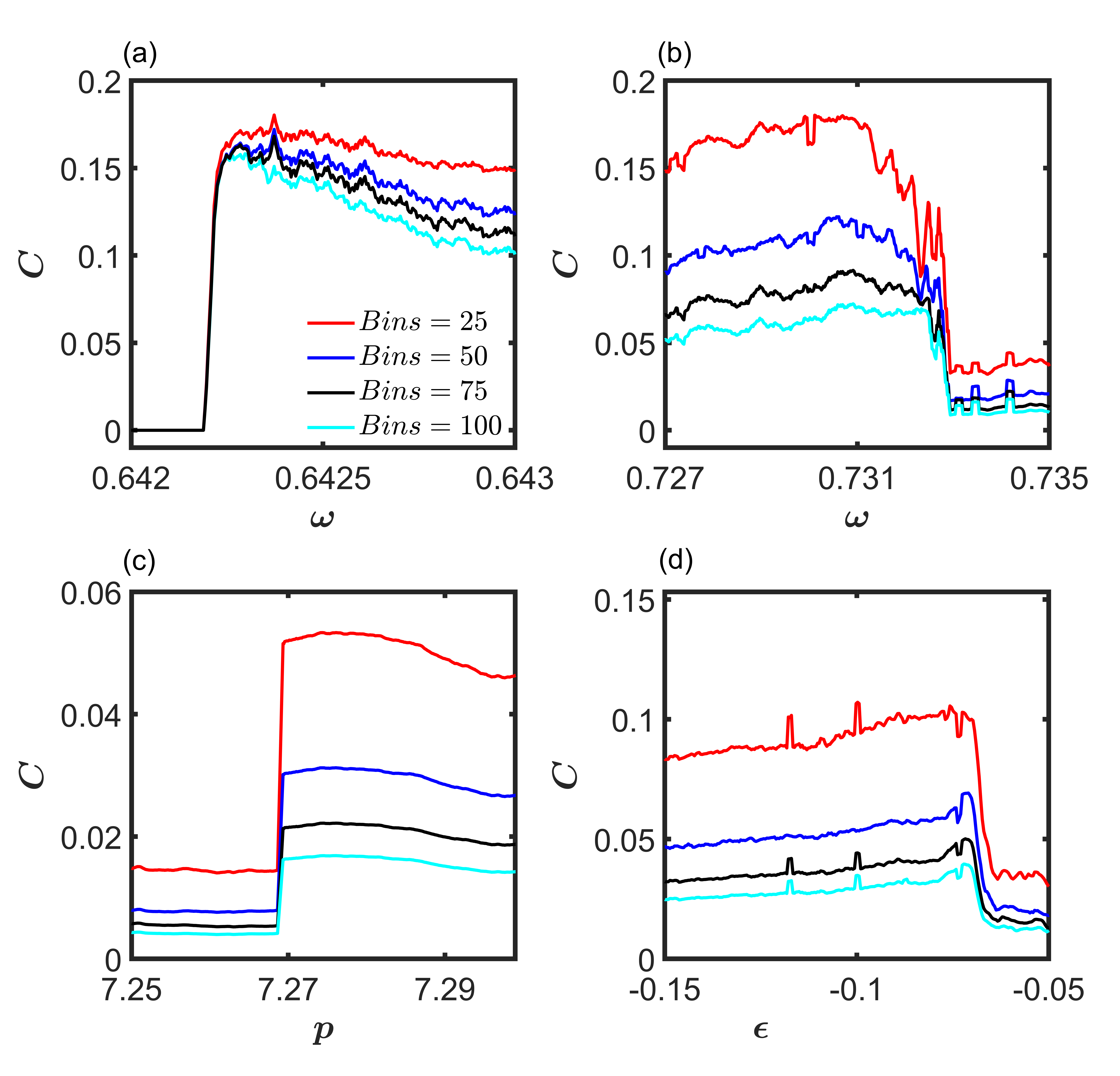}
\caption{Dependence of $C$ and its transition points on bin size during PM intermittency (a) and interior crisis (b) in Li\'enard system, during interior crisis in (c) Ikeda map and (d) coupled HR model. Different  bin sizes, $m$=[25, 50, 75, 100]  }
    \label{fig5}
    \end{figure}\\


\par {\bf Acknowledgments:}
D. Das would like to thank Tapas Kumar Pal for
the helpful discussions. D. Das thanks the University Grants Commission (UGC), India, for providing financial support (NTA Ref. No. 211610125989). D. Das, S.K.D., T.K. and A.D. acknowledge financial support by the National
Science Centre, Poland, OPUS Programs (Projects No.
2018/29/B/ST8/00457, and 2021/43/B/ST8/00641).
D.G. is also supported by DST-SERB Core Research Grant (Project No. CRG/2021/005894). 
S.K.D. dedicated to his Late father, Manik C. Dana, on his 100th years of birth anniversary.

\section*{AUTHOR DECLARATIONS}
\section*{Conflict of Interest}
The authors have no conflicts to disclose.

\section*{DATA AVAILABILITY}
The data that support the findings of this study are available from the corresponding author upon reasonable request.


\section*{References}
%

\end{document}